\documentclass[superscriptaddress,twocolumn,pra,10pt,aps,showpacs,amsmath,amssymb,longbibliography]{revtex4-1}
\usepackage{dcolumn}
\usepackage{graphicx}
\usepackage{amsfonts} 
\usepackage{epstopdf}
\usepackage{color}
\usepackage{here}
\usepackage{hyperref}
\usepackage{circuitikz}

\graphicspath{{./img/}}

\begin{document}
\title{Phonon-Josephson resonances in atomtronic circuits}
\date{\today}
\author{Y.M. Bidasyuk}
\email{Yuriy.Bidasyuk@ptb.de}
\affiliation{
Physikalisch-Technische Bundesanstalt, Bundesallee 100, D-38116 Braunschweig, Germany
}
\author{O.O.Prikhodko}
\affiliation{
Department of Physics, Taras Shevchenko National University of Kyiv, Volodymyrska Street 64/13, Kyiv 01601, Ukraine
}
\author{M. Weyrauch}
\affiliation{
Physikalisch-Technische Bundesanstalt, Bundesallee 100, D-38116 Braunschweig, Germany
}
\begin{abstract}
We study the resonant excitation of sound modes from Josephson oscillations in Bose-Einstein condensates.
From the simulations for various setups using the Gross-Piaevski mean field equations and Josephson equations we observe additional tunneling currents induced by resonant phonons. The proposed experiment may be used for spectroscopy of phonons as well as other low-energy collective excitations in Bose-Einstein condensates.
We also argue that the observed effect may mask the observation of Shapiro resonances if not carefully controlled.
\end{abstract}
\maketitle
\section{Introduction}
Superconducting Josephson junctions provide high-precision frequency-to-voltage  conversion known as the Shapiro effect~\cite{PhysRevLett.11.80,barone}.
In fact, the modern standard of the volt is based on arrays of Josephson junctions connected to a microwave field with frequency controlled by an atomic frequency standard~\cite{grandstrand:2004,0957-0233-23-12-124002,hamilton2000josephson}.
Moreover, as an essential building block of superconducting quantum interference devices (SQUIDs) Josephson junctions provide  magnetic-flux-to-voltage conversion enabling high-precision magnetic-flux measurements~\cite{squidbook}.

Recently, advanced techniques in optical trapping and control initiated  studies of Josephson effects in atomic Bose-Einstein condensates (BECs).
Narrow barriers created by blue-detuned laser beams operate as Josephson junctions with highly adjustable properties.
In this way one can produce simple two-well traps~\cite{levy2007ac,PhysRevLett.95.010402}, atomic SQUIDs with one or two junctions~\cite{PhysRevLett.111.205301,PhysRevLett.113.045305,eckel2014hysteresis,1367-2630-17-12-125012}, or even arrays of Josephson junctions in optical lattices~\cite{RevModPhys.78.179,PhysRevLett.100.040404}.
Extensive theoretical and numerical studies investigate various dynamics of BECs with such barriers and determine the requirements for barriers to operate in the Josephson regime~\cite{RevModPhys.73.307,PhysRevA.59.620}. Moreover, simple mathematical models based on the two-mode approximation were develloped and supplemented with corrections for non-linear interactions and asymmetric trap configurations in order to understand atomtronic Josephson physics~\cite{PhysRevA.90.043610}.
Various experimental and theoretical results demonstrate how the quantum nature of the Josephson effect and the existence of critical tunneling currents lead to the creation of a chemical potential difference in two-well systems \cite{levy2007ac,PhysRevLett.95.010402,PhysRevLett.111.205301,PhysRevLett.113.045305}. On the microscopic level vortex-mediated phase slips were observed and used to study such phenomena as persistent currents and quantum hysteresis \cite{PhysRevLett.106.130401,eckel2014hysteresis,abad2015phase,PhysRevA.91.033607,
PhysRevA.91.023607,yakimenko2014generation}.
The existence of critical currents and phase slips in the over-critical region are cornerstones of Josephson physics in BECs.

Nevertheless, some aspects of Josephson physics in atomic BEC still remain little explored. One of them is the Shapiro effect, i.e. the use of atomtronic Josephson devices as precise frequency-to-chemical potential converters. 
There are several
theoretical investigations of the Shapiro effect in a trapped BEC with a single Josephson barrier~\cite{KohleSols03,PhysRevLett.95.200401,1367-2630-13-6-065026}. While such an effect has been observed in optical lattices~\cite{PhysRevLett.100.040404},
experimental observations of Shapiro effects in single- or double-junction setups are lacking.
This may be due to the fact  that coupling of Josephson oscillations to other excitations with similar frequencies masks the observation of Shapiro resonances. The existence of such coupling effects is well known from superconducting
Josephson systems~\cite{PhysRev.138.A744,Maksimov1999391,PhysRevLett.79.737} as well as from  superfluid $^3$He~\cite{PhysRevLett.81.1247}.

In the present work we study resonant coupling between Josephson oscillations and phonon modes in atomic BECs. To this end we propose to investigate an experiment on the basis of a toroidal trap used to realize Josephson junctions for atomtronic SQUIDs~\cite{PhysRevLett.111.205301}.
The trap is divided into two parts by two optical Josephson barriers, and these two parts are initially populated with condensates of different number density.
If this imbalance is small  one expects imbalance oscillations around zero mean (often called plasma oscillations in analogy to superconductors).
For larger imbalances one expects Josephson oscillations of the number density imbalance around a non-zero mean, which is also known as macroscopic quantum self-trapping (MQST).
However, if phonon modes can be resonantly exited by the Josephson alternating current, then this coupling will provide a dissipation channel.
This should happen if the phonon frequencies of the trap match the Josephson current frequency.
We therefore expect to observe some characteristic resonant response from the system in this region, which can be observed in the evolution of the population imbalance or the chemical potential difference between two wells.
In this paper we propose an experimentally feasible protocol to observe the dynamical picture outlined above.
The simulations of such an experiment are made using the three-dimensional time-dependent Gross-Pitaevskii equation (GPE). The results are supported by a simplified model based on the Josephson equations.

\section{Theoretical setup}

We consider the toroidal harmonic trap used in Ref.~\cite{PhysRevLett.111.205301} for the experimental realization of atomtronic Josephson junctions. It is descibed by the potential
\begin{equation}
V_{\rm trap}(\mathbf{r}) = \frac12 M \omega_z^2 z^2 + \frac12 M \omega_0^2 (r_\perp - r_0)^2,
\end{equation}
with $r_\perp=\sqrt{x^2+y^2}$, ring radius $r_0=4~\mu$m, and frequencies $\omega_z/2\pi=300$~Hz and $\omega_0/2\pi=570$~Hz. 
In comparison to other toroidal traps \cite{PhysRevLett.106.130401,eckel2014hysteresis,PhysRevA.91.013602,PhysRevLett.113.045305,1367-2630-17-12-125012}, this trap is considerably elongated in the $z$ direction. 
The advantage of this trap for our purposes is that the nucleation of vortices is suppressed (see also Ref.~\cite{1367-2630-13-4-043008}). 
The formation of annular vortices may significantly distort the dynamics we intend to observe. Such a trap configuration requires three-dimensional simulations of the condensate dynamics.

The trap is subdivided by static barriers into two parts created by scanning blue-detuned laser beams across the trap annulus. They can be represented by repulsive potentials $V_1$ and $V_2$ , which are homogeneous in the radial direction and have a Gaussian shape in the tangential direction \cite{PhysRevA.91.033607},
\begin{equation}
V_{i}(\textbf{r}_\perp)=U_{i}\Theta(\textbf{r}_\perp\cdot \textbf{n}_{i})e^{-[\textbf{r}_\perp \times \textbf{n}_{i}]^2 / d^2}.
\end{equation}
The unit vectors $\textbf{n}_{i}$ ($i=1,2$) point radially into the directions of the barriers, $\textbf{r}_\perp=\left(x,y\right)$ is a vector in the $x y$ plane, $\Theta$ is the Heaviside Theta function, and $d=1.2~\mu$m is the $1/e$ half-width of the barrier (corresponding to the $2~\mu$m full width at half maximum in the experiment).
The experimental barrier height is reported to be $44~$nK which corresponds to $U_1/h=U_2/h=917.3~$Hz. As reported in \cite{PhysRevLett.111.205301} and supported by the results discussed below, such barriers operate in the Josephson regime.

The initial population imbalance between the two parts of the trap is produced by an additional tilt potential (not present in the experimental setup of Ref.~\cite{PhysRevLett.111.205301}).
The total external potential $V$ in which the BEC moves therefore consists of the trapping potential, the barrier potential $V_b=V_1+V_2$, and the time-dependent tilt potential,
\begin{equation}
V(\mathbf{r},t) = V_{\rm trap}(\mathbf{r}) + V_b(\mathbf{r}) + V_{\rm tilt}(\mathbf{r},t).
\end{equation}
The initial tilt is linearly switched off within $\tau=0.01~$s after starting the simulation of the BEC dynamics,
\begin{equation}
V_{\rm tilt}(\mathbf{r},t) = \left\{\begin{aligned}
U_0\,x(1-t/\tau), &\qquad t\le\tau \\
0, &\qquad t>\tau.
\end{aligned}\right.
\end{equation}
Then the evolution of the trapped condensate is observed for an additional $0.5~$s.
\begin{figure}[htbp]
\includegraphics[width=\linewidth]{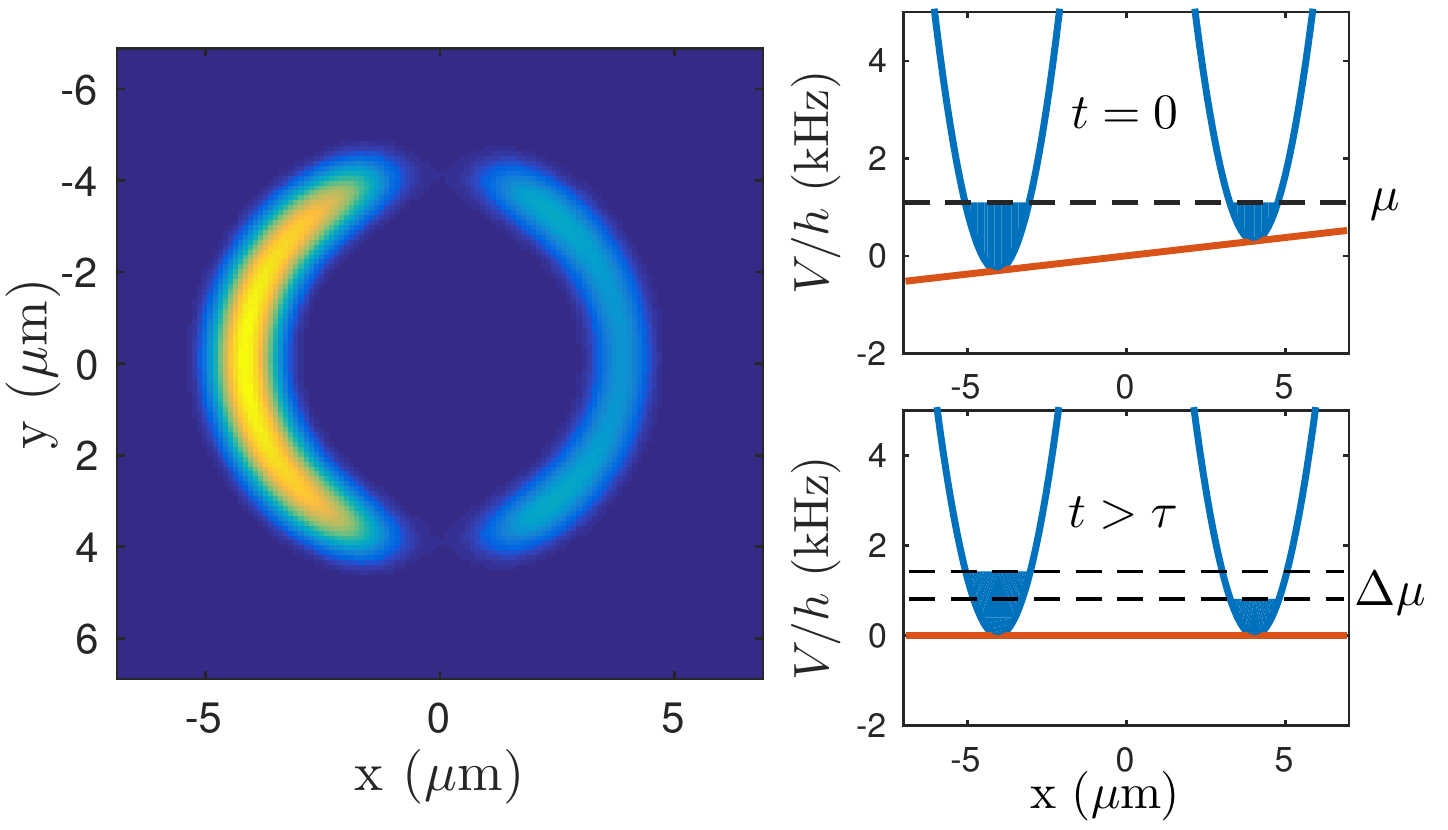}
\caption{
Shown on the left is the density distribution of the simulated BEC cloud in $xy$-plane in the tilted toroidal trap with two barriers.
Shown on the right is the creation of the chemical potential difference in the proposed experimental protocol. The figures show the trapping potential along the $x$-axis (blue lines), tilt potential (straight red lines) and Thomas-Fermi populations in each part (blue shaded regions) at two different times.
}\label{fig:config}
\end{figure}

The proposed trap configuration is shown in Fig.~\ref{fig:config}, which illustrates how the tilt potential produces the initial population imbalance between the two parts of the ring.
When the tilt potential is switched off, this initial population imbalance leads to a chemical potential difference $\Delta\mu$.

The BEC is described by a macroscopic wave function $\Psi(\mathbf{r},t)$, which obeys the time-dependent GPE
\begin{equation}\label{GP}
i \hbar \frac{\partial \Psi}{\partial t} = -\frac{\hbar^2}{2 M} \nabla^2 \Psi + (V + g |\Psi|^2) \Psi.
\end{equation}
The wave function is normalized to the total number of atoms
$
 N_T=\int|\Psi|^2d\textbf{r},
$
which is chosen to be $N_T=5000$ unless explicitly stated otherwise. The nonlinear coupling $ g = 4 \pi \hbar^2 a_s/{M}$ is given in terms of the mass $M$ of the $^{87}$Rb atom  and its $s$-wave scattering length $a_s$.

The populations in each part of the trap can be obtained by integration over each nonoverlapping part $W_i$  separately,
$
N_{i}(t) = \int_{W_{i}} |\Psi(\mathbf{r},t)|^2\,d\mathbf{r},
$
with $i=L$ (left) or $i=R$ (right) and $N_T= N_L(t)+N_R(t)$. The total particle number $N_T$ is an integral of motion of the GP equation (\ref{GP}) and therefore time independent.
The relative population imbalance is then given by
$$
Z(t)=\frac{N_L(t)-N_R(t)}{N_T}.
$$
Analogously, we obtain the local chemical potentials $\mu_L(t)$ and  $\mu_R(t)$ of each part of the trapped condensate as well as the chemical potential difference $\Delta\mu(t) = \mu_L(t) - \mu_R(t)$ from
\begin{equation}\label{GPmu}
\mu_i = \frac{1}{N_i} \int_{W_i}  \left[ -\frac{\hbar^2}{2 M} | \nabla \Psi |^2 + V |\Psi|^2 + g |\Psi|^4 \right] d\textbf{r}.
\end{equation}

In terms of these quantities one may describe the dynamics of the BEC by a two-mode approximation to the GP equation~\cite{PhysRevA.59.620,PhysRevA.90.043610} often termed Josephson equations,
\begin{eqnarray}\label{JEFs}
\dot Z(t) &=& -\omega_J\sqrt{1-Z(t)^2} \sin[\phi(t)],\nonumber\\
\dot \phi(t) &=&\Delta \mu(t)/\hbar,
\end{eqnarray}
where $\phi$ is the phase difference between the two parts of the condensate.
This set of equations is closed by the relation $\Delta\mu(t) = \hbar\omega_C Z(t)$.

The evolution of the condensate is uniquely determined by the initial population imbalance $Z(0)=Z_0$, the capacitive energy $E_C = 2\hbar\omega_C/N_T$, and the Josephson critical current $\omega_J$, which is related to the Josephson coupling energy $E_J = N_T\hbar\omega_J/2$.
Values for these quantities can be estimated from stationary solutions~\cite{PhysRevA.90.043610}, which we calculate using imaginary-time evolution of the Eq.~(\ref{GP}) with a static tilt potential.
We find $\omega_J/2\pi = 0.68$~Hz and $\omega_C/2\pi=851$~Hz as well as the initial population imbalance $Z_0$ for each value of $U_0$.
The ratio $\omega_C/\omega_J$ is well inside the region $1 \ll \omega_C/\omega_J \ll (N_T/2)^2$, supporting that the barriers operate in the Josephson regime~\cite{RevModPhys.73.307}.
Since the ratio between the barrier hight and the total chemical potential is about  $U_{1,2}/\mu \approx 0.8$, the barriers can be considered to be Josephson weak links.
Our GPE simulations clearly support the expected linear relation between $Z$ and $\Delta\mu$ which allows us to use the relative population imbalance as a measure of the chemical potential difference.
In the presentation of our results below we will use either of these two quantities interchangeably, whichever is more illustrative.
From the simulations we also confirm the linear relation between $\dot Z$ and $\sin(\phi)$, another indication that the barriers operate in the Josephson regime.

\subsection*{Excitations in a trapped condensate}

\begin{figure}[htbp]
\includegraphics[width=\linewidth]{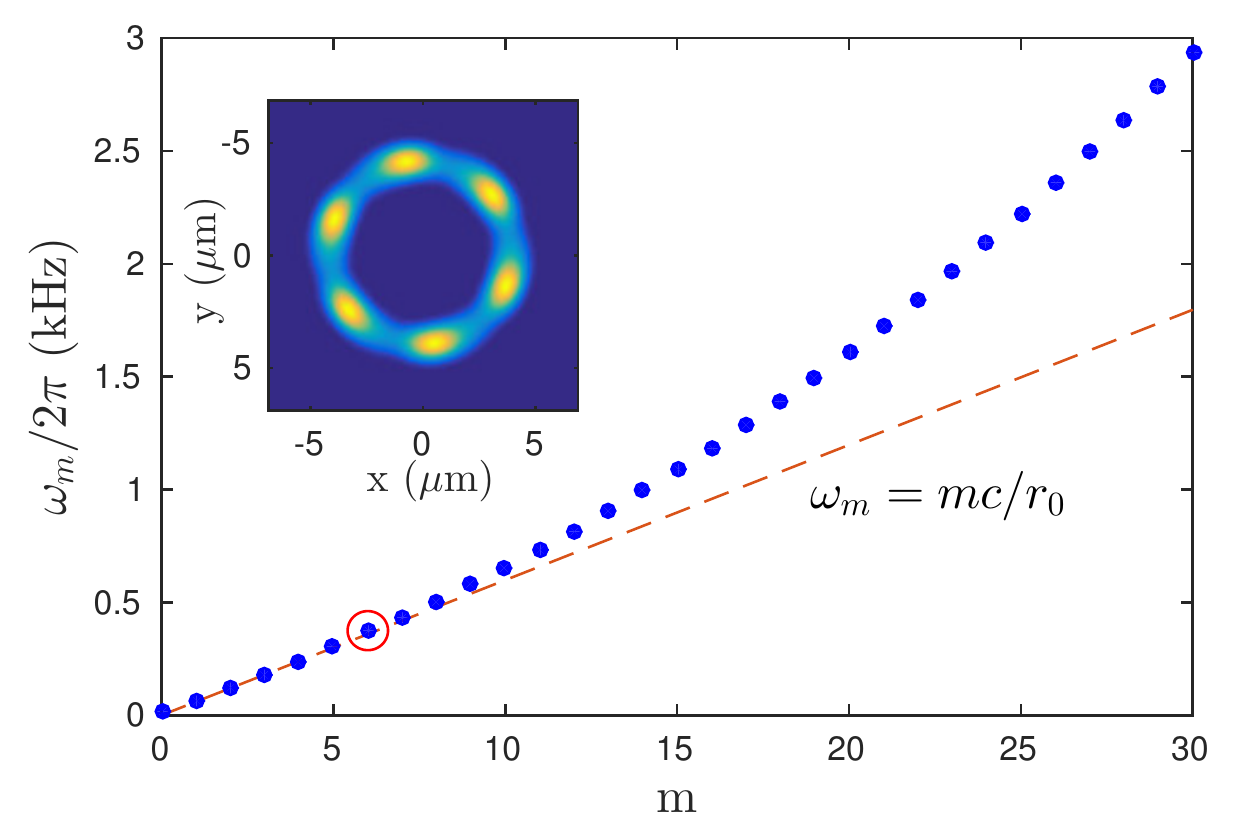}
\caption{Lowest branch of the BdG spectrum for elementary excitations of the condensate in a symmetric ring trap.
The dashed line represents the sound frequencies estimated from the average density.
The inset shows the particle density $|\Psi|^2$ in the $xy$ plane with 25\% of the population in the $m=6$ mode (marked with a red circle).
}\label{fig:pjres_bdg}
\end{figure}

In order to identify the elementary excitations in the trapped BEC system under consideration we employ the Bogolyubov-de-Gennes (BdG) formalism (see e.g.~\cite{1367-2630-17-12-125012,PhysRevA.86.011602,PhysRevA.86.011604}).
We first consider a condensate without tilt and barrier potentials and write the cylindrically symmetric BEC wave function in the form
\begin{equation}
\Psi(\mathbf{r},t) = e^{-i\mu t} \left[ \Psi_0(r_\perp,z) + \delta \Psi(\mathbf{r}_\perp,z,t) \right],
\end{equation}
where $\Psi_0$ is the stationary solution of (\ref{GP}) with chemical potential $\mu$. The perturbation
\begin{equation}
\delta \Psi(\mathbf{r}_\perp,z,t) = u_m (r_\perp,z) e^{-i(\omega t - m \theta)} + v_m^* (r_\perp,z) e^{i(\omega t - m \theta)}
\end{equation}
is characterized by a well-defined azimuthal quantum number $m$ due to the cylindrical symmetry.
We insert this ansatz into Eq.~(\ref{GP}) and linearize it with respect to $u_m$ and $v_m$. From the resulting BdG equations one finds for each $m$ a set of eigenvalues $\omega$.
The smallest eigenvalue for each $m$ determines the lowest branch $\omega_m$ of the excitation spectrum, which is shown in Fig.~\ref{fig:pjres_bdg}.

For small quantum numbers $m$ the dispersion law is linear, which suggests  that the excitations are soundlike modes (phonons).
This part of the excitation spectrum  may be expressed in terms of the speed of sound $c$ and the ring radius $r_0$, $\omega_m = mc/r_0$. Alternatively, the speed of sound  can be related to the average density $\tilde n$ of the condensate, $c=\sqrt{g\tilde n/M}$ \cite{PhysRevA.57.518}.
Estimating the average density as one-half of the peak density of the stationary state, we obtain values for $c$
in very good agreement with those obtained from the BdG spectrum as illustrated by the dashed red line in Fig.~\ref{fig:pjres_bdg}.

In the presence of barriers phonon excitations become standing waves localized in either of two parts of the condensate with nodes at the barrier's positions.
Both the average density and the angular extent of the two condensate parts $L$ or $R$ can be different, therefore, in principle one needs to consider two different phonon frequencies
$
\Omega_{i} = {2\pi\, c_{\rm i}}/({\alpha_{i}\, r_0}),
$
with  $i=\{L,R\}$ denoting the left and right parts, respectively, $\alpha_{i}$ is the angular size of each part, and $c_i$ is the speed of sound in each part.

\section{Phonon-Josephson resonances}\label{sec:mainres}

The frequency of Josephson oscillations is related to the  chemical potential difference~[see Eq.~(\ref{JEFs}) and Ref.~\cite{levy2007ac}].
For a system operating in the Josephson regime we therefore expect to observe resonant coupling between Josephson oscillations and phonon modes under the conditions
\begin{equation}\label{eq:res_cond}
\Delta\mu / \hbar = m \Omega_{L,R}, \qquad m=1,2,\ldots .
\end{equation}
To verify this expectation we perform a number of simulations based on the GPE~(\ref{GP}).

\begin{figure}[htbp]
\includegraphics[width=\linewidth]{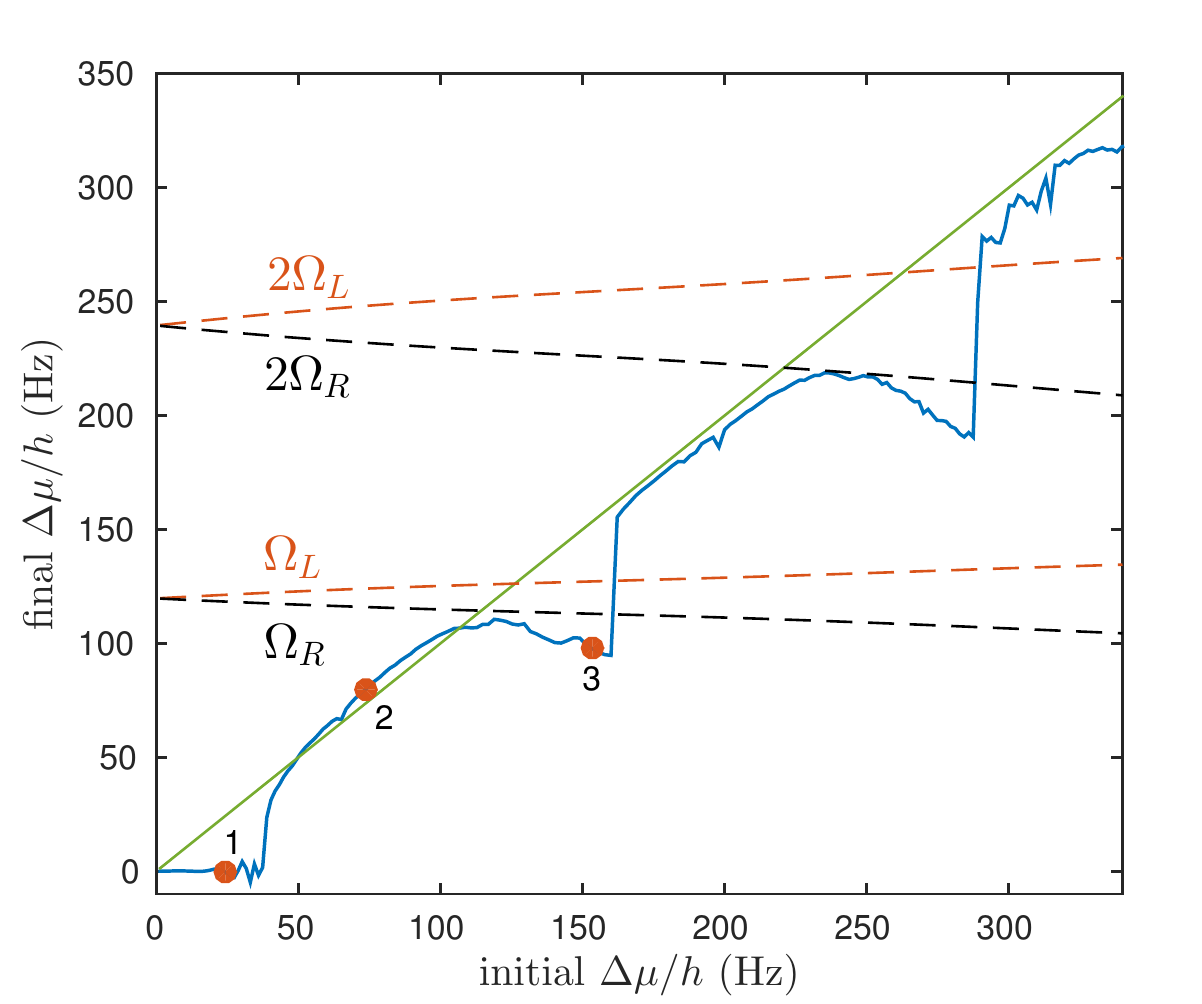}
\caption{
Demonstration of phonon-Josephson resonances.
The solid blue line shows the final chemical potential difference as a function of initial chemical potential difference.
Dashed lines represent estimated resonance positions for phonon modes.
The straight green line is a guide to the eye marking equal initial and final values of $\Delta\mu$.
}\label{fig:fiske}
\end{figure}

We begin with a symmetric barrier configuration, i.e., both barriers are located along the $y$ axis as shown in Fig.~\ref{fig:config}.
The angular extent of both parts is $\alpha_L = \alpha_R = \pi$.
We simulate the dynamics of the GPE (\ref{GP}) using a split-step Fourier transform method.
Each simulation starts with a different initial tilt $U_0$ providing us with different values of initial chemical potential difference $\Delta\mu(t=\tau)$. We then measure the final chemical potential difference as the average value over the last 0.2 s of the evolution in order to eliminate the effects of high-frequency Josephson or plasma oscillations.
Such a series of simulations allows us to study the final $\Delta\mu$  as a function of the initial $\Delta\mu$ and identify those regions where they are different.
The results of such a series of simulations are presented in Fig.~\ref{fig:fiske}.

\begin{figure}[htbp]
\includegraphics[width=\linewidth]{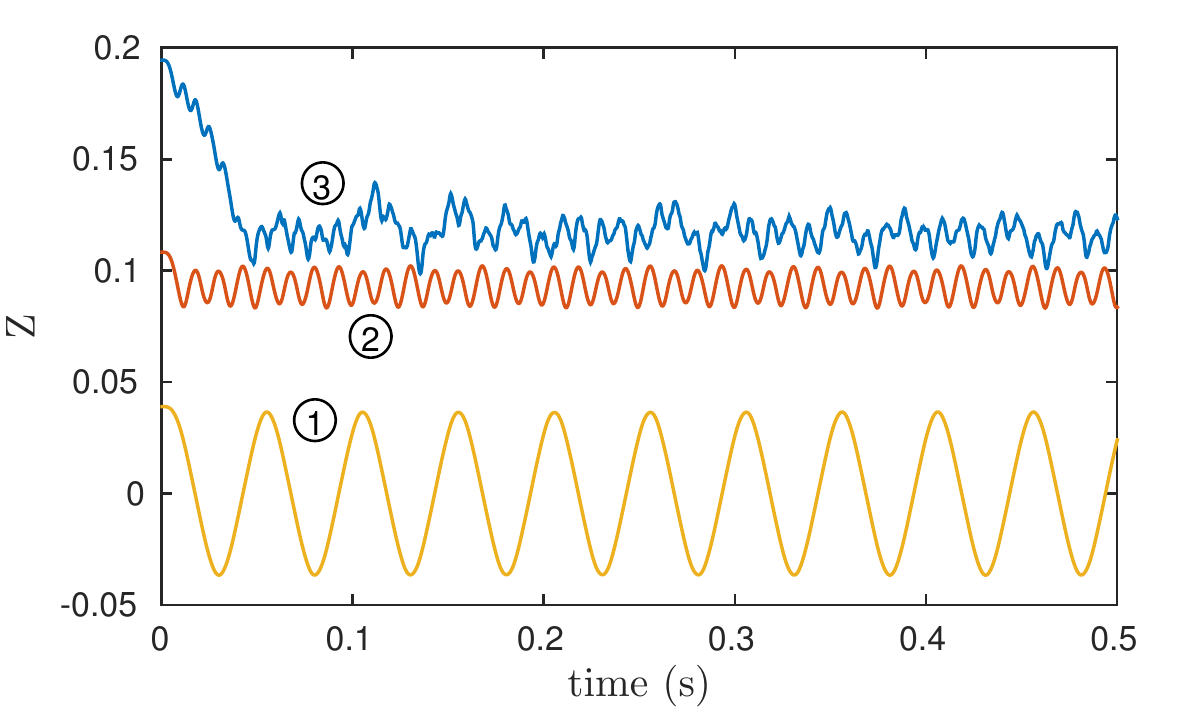}
\caption{
(Color online) Time evolution of the relative population imbalance for the system in the regimes of plasma oscillations (1), MQST (2), and phonon-Josephson resonance (3).
}\label{fig:phonon_zt}
\end{figure}

There are three different regimes observed in the system's evolution depending on the initial imbalance. These regimes are marked as 1, 2, and 3 in Fig.~\ref{fig:fiske}. Corresponding curves in Fig.~\ref{fig:phonon_zt} show the actual time evolution of the population imbalance.
For small values of the initial population imbalance the system is in the plasma oscillation regime, and the time average of the population imbalance and the chemical potential difference are zero.
With larger initial population imbalance the system switches into the self-trapped regime (MQST), and according to the Josephson dynamics the average population imbalance should remain close to its initial value.
This is indeed the case everywhere except for two distinct regions, where the final population imbalance is  reduced due to the resonant dissipation into phonon excitations.
As expected the resonance occurs in regions where the Josephson frequency is close to the frequency of a phonon mode.
The frequencies $\Omega_L$ and $\Omega_R$ of the phonon modes in Fig.~\ref{fig:fiske} are estimated using the density distributions of each initial stationary state.
These frequencies are equal for zero initial tilt and differ for growing population imbalance. We notice that the resonance peak is observed at slightly larger $\Delta\mu$ than expected from Eq.~(\ref{eq:res_cond}).

\begin{figure}[htbp]
\includegraphics[width=\linewidth]{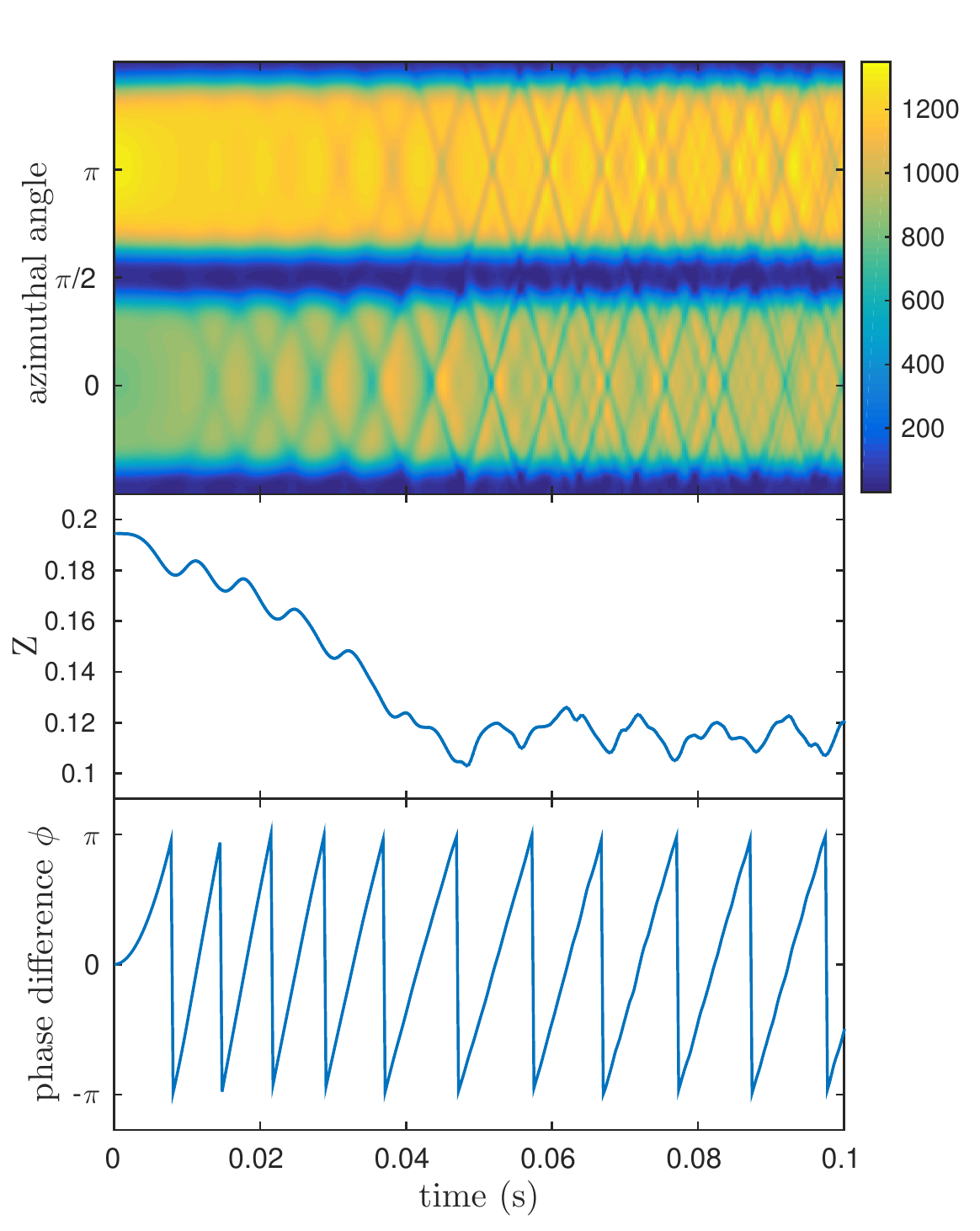}
\caption{
Generation of a phonon mode by Josephson oscillations.
The top panel shows the azimuthal distribution of the condensate density $n=|\Psi|^2$ as it changes with time (dark blue regions correspond to the barriers positioned at $\pi/2$ and $-\pi/2$).
The middle and bottom panels show the evolution of population imbalance $Z$ and phase difference $\phi$ within the same time period.
}\label{fig:phonon_az_dens}
\end{figure}

In the resonance region the excitation of phonon modes can be clearly seen in the density distribution of the condensate (see Fig.~\ref{fig:phonon_az_dens}).
From the time dependence of the relative population imbalance $Z(t)$ one can see that the generation of phonon modes leads to a decrease of the population imbalance.
This can be understood as a transfer of energy stored in the chemical potential difference to the phonon mode through resonant coupling.
Such phonon-assisted dissipation quickly drives the system out of resonance (within the first $\sim 0.04$~s), lowering the Josephson oscillation frequency. 
This breaks the coupling and the population imbalance oscillates around its new average value.
When the system is out of resonance, phonon modes are not excited and the system again shows  MQST dynamics.

The latter statement is supported by the trace of the phase difference shown in Fig.~\ref{fig:phonon_az_dens}.
Between the phase slips the phase difference grows linearly, which is typical for the MQST regime.
Its slope and consequently the frequency of phase slips changes after the population imbalance is reduced by
the resonant coupling to phonons, but otherwise the Josephson dynamics is not affected.

\begin{figure}[htbp]
\includegraphics[width=0.95\linewidth]{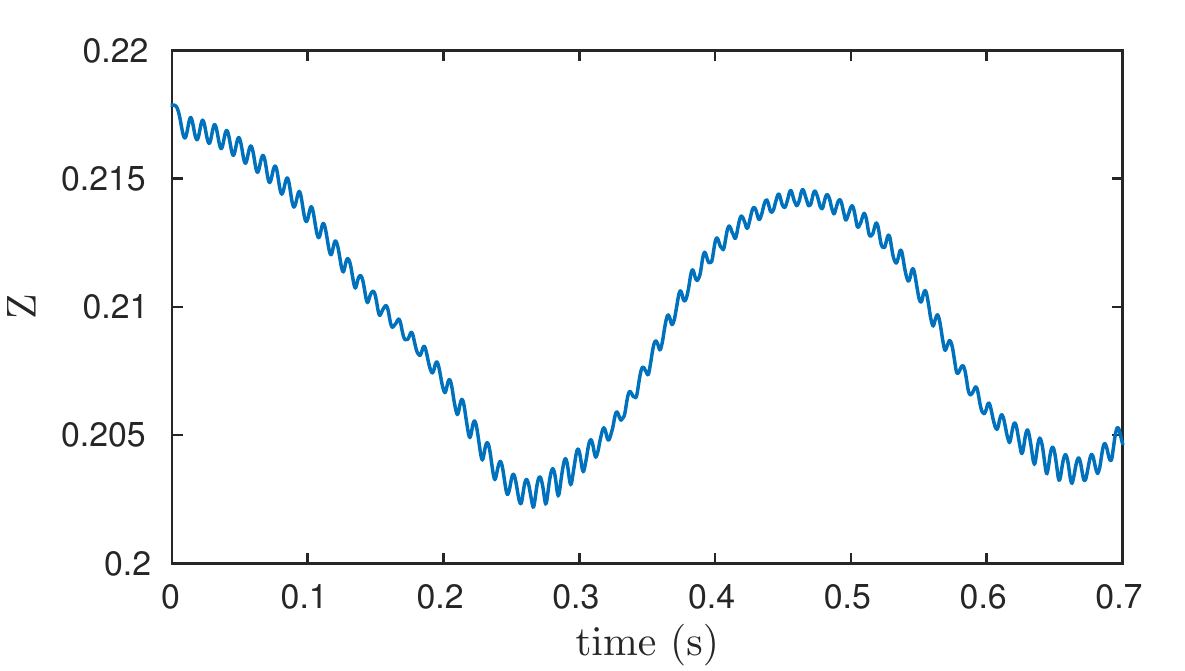}
\caption{Beats of the population imbalance for the system with $N_T=2000$ particles in near-resonance conditions. Slow oscillations result from the interference of the phonon mode and Josephson oscillations (fast oscillations of $Z$) with similar frequencies.
}\label{fig:pjres_beats}
\end{figure}

We now study the Josephson dynamics in the proposed trap with a significantly reduced amplitude of the Josephson oscillations. This can be achieved in various ways, e.g. by a modification of the barrier parameters or a reduction of the total number of particles.
By lowering this amplitude we eventually observe a situation when the system is not driven out of resonance. Instead the energy, which was resonantly transferred to the phonon mode, is transferred back into the chemical potential difference, i.e. Josephson oscillations.
As an example we consider a BEC in the same trap as before with only $N_T=2000$ atoms.
Then one finds $\omega_J/2\pi=0.05$~Hz and $\omega_C/2\pi=627$~Hz and the system is still in the Josephson regime.
The amplitude of the Josephson oscillations is reduced by about one order of magnitude.
In this case one does {\it not} observe any noticeable drop of final chemical potential difference in the resonance regions.
The system is not driven out of the resonance conditions and the energy transferred from the Josephson oscillations to the phonon mode can be resonantly transferred back into Josephson oscillations, reducing the energy of the phonon mode and increasing the chemical potential difference.
The system behaves like coupled oscillators showing characteristic beats as shown in Fig.~\ref{fig:pjres_beats}.

\subsection*{Equivalent description using Josephson equations}

In this subsection we describe the dynamics observed in our Gross-Pitaevskii simulations by equivalent Josephson equations~(\ref{JEFs}).
The system described by the Josephson equations can be considered as an electric circuit consisting of a capacitor with capacitance $1/\hbar\omega_C$ initially charged to  $Z_0$ and connected to an ideal Josephson junction. The second Josephson equation represents a second Kirchoff law for such a circuit.

\begin{figure}[htb]
\includegraphics[width=\linewidth]{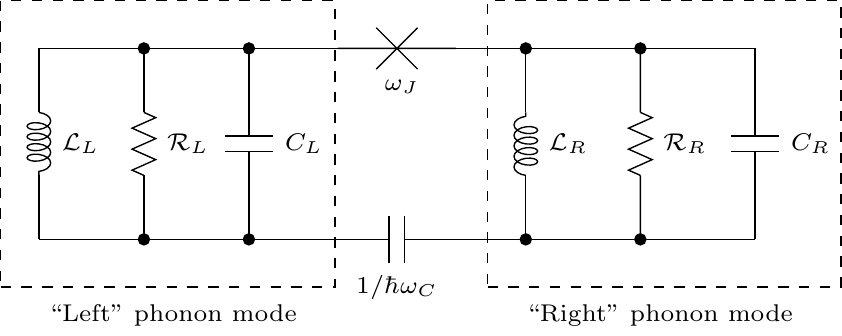}
\caption{Equivalent electric circuit for the toroidal condensate with two weak links. It consists of the Josephson junction, the initially charged capacitor with charge $Z$, which represents the total population imbalance, and two $RLC$ circuits, which represent phonon modes in two parts of the condensate.
}\label{fig:circuit}
\end{figure}

Let us now consider a more complicated circuit (Fig.~\ref{fig:circuit}) that contains in addition two parallel $RLC$ circuits connected in series to the capacitor and  Josephson junction. For this case the second Kirchoff law must additionally include voltage drops on these two $RLC$ circuits. This setup results in the set of equations
\begin{eqnarray}\label{JEFs2}
\dot Z(t) &=& -\omega_J\sqrt{1-Z(t)^2} \sin[\phi(t)]\nonumber\\
\dot \phi(t) &=&\omega_C Z + \frac{u_L}{\hbar} + \frac{u_R}{\hbar}
\end{eqnarray}
with additional equations for $u_L$ and $u_R$:
\begin{equation}\label{eq:kirchhof}
\frac{1}{\mathcal{R}_i} u_i + C_i \dot u_i + \frac{1}{\mathcal{L}_i} \int\limits_0^t u_i(\tau) d\tau = \dot Z(t), \quad i=\{L,R\},
\end{equation}
where $\dot Z(t)$ represents the total current through the $RLC$ circuit and three terms in the left-hand side are partial currents through each branch of the circuit.
The inductance and capacitance in each circuit are chosen to match the calculated phonon frequencies $1/\sqrt{\mathcal{L}_{i}C_{i}}=\Omega_{i}$.
We also consider equal inductances $\mathcal{L}_L=\mathcal{L}_R=\mathcal{L}$ and resistances $\mathcal{R}_L=\mathcal{R}_R=\mathcal{R}$ for simplicity, which means that we have only two free parameters in the model, $\mathcal{L}$ and $\mathcal{R}$. They are obtained by fitting the GPE results.
Such fitting gives $\mathcal{R}/\hbar=8000$ and $\mathcal{L}/\hbar=61$~ms, which results in a damping factor of the $RLC$ circuit $\zeta = \sqrt{\mathcal{L}/C}/2\mathcal{R} \approx 4 \cdot 10^{-3}$.

\begin{figure}[htbp]
\includegraphics[width=\linewidth]{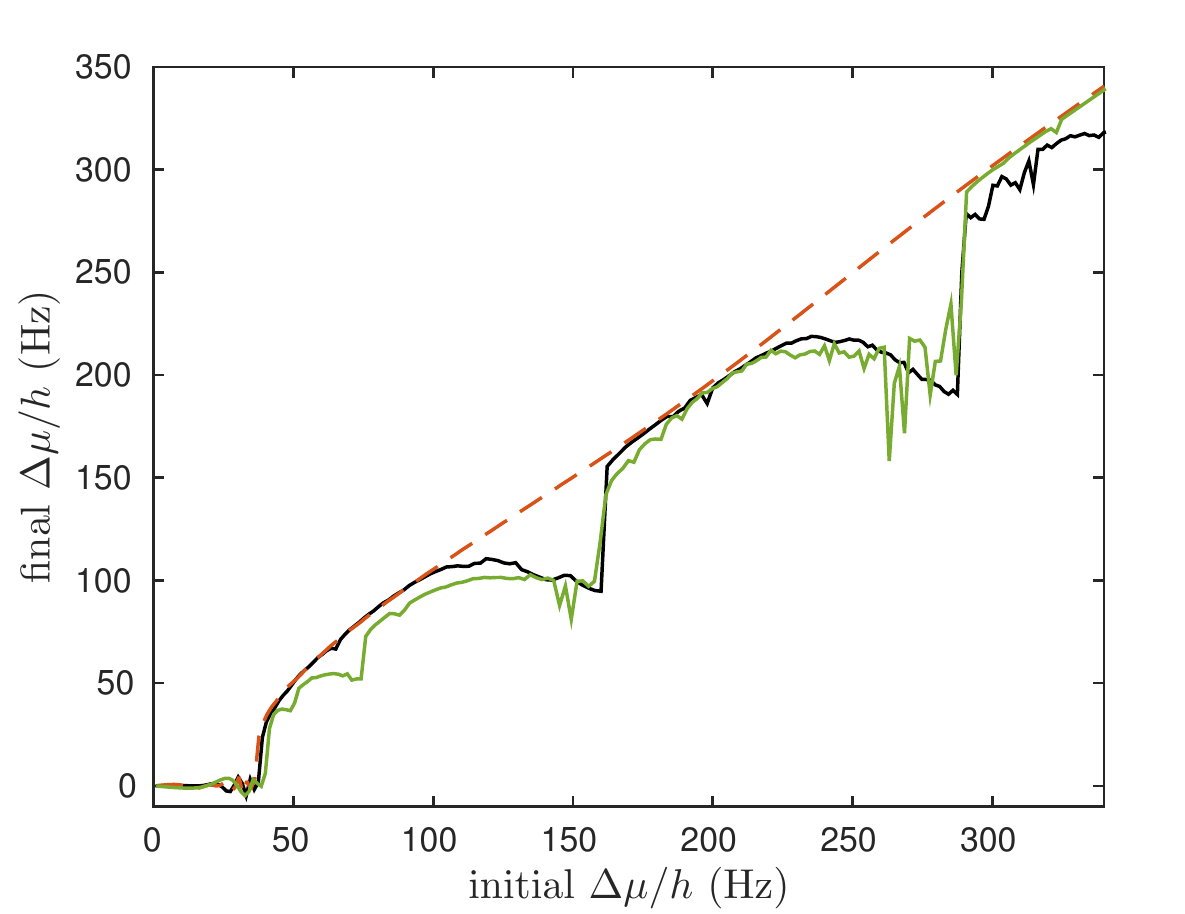}
\caption{(Color online) Final chemical potential difference as a function of initial chemical potential difference obtained from the GPE calculation (black line) and Josephson equations with (green line) and without (red dashed line) additional $RLC$ circuits representing phonon modes.
}\label{fig:pjres_z_eqcirc}
\end{figure}

From the results in Fig.~\ref{fig:pjres_z_eqcirc} one can see that such an equivalent resonating circuit correctly captures the qualitative resonant picture, and even quantitative agreement can be achieved except in the region of high population imbalance, where other excitations become relevant for the condensate dynamics. 
It is also worth noticing that a small subharmonic resonance at $\Delta\mu / \hbar = \Omega/2$, which is barely visible in the results of GPE dynamics, is much more pronounced in the equivalent circuit.

\section{Non-symmetric traps}

In this section we describe simulations of phonon-Josephson resonances in non-symmetric traps either
by choosing the barriers that divide the trap into parts of unequal size, or by placing
barriers with unequal properties into the trap.

\subsection{Asymmetric positions of the barriers}

\begin{figure}[htbp]
\includegraphics[width=\linewidth]{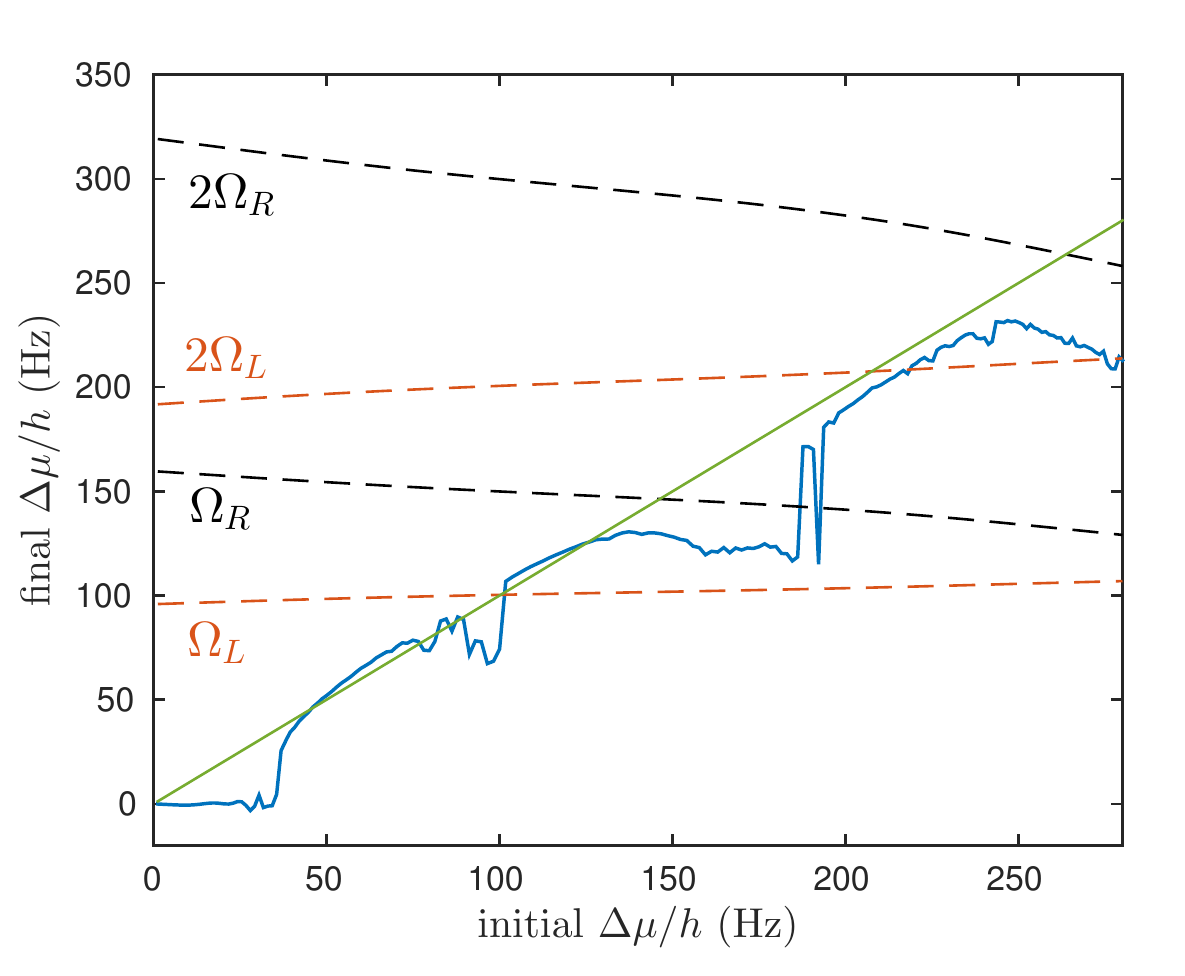}
\includegraphics[width=\linewidth]{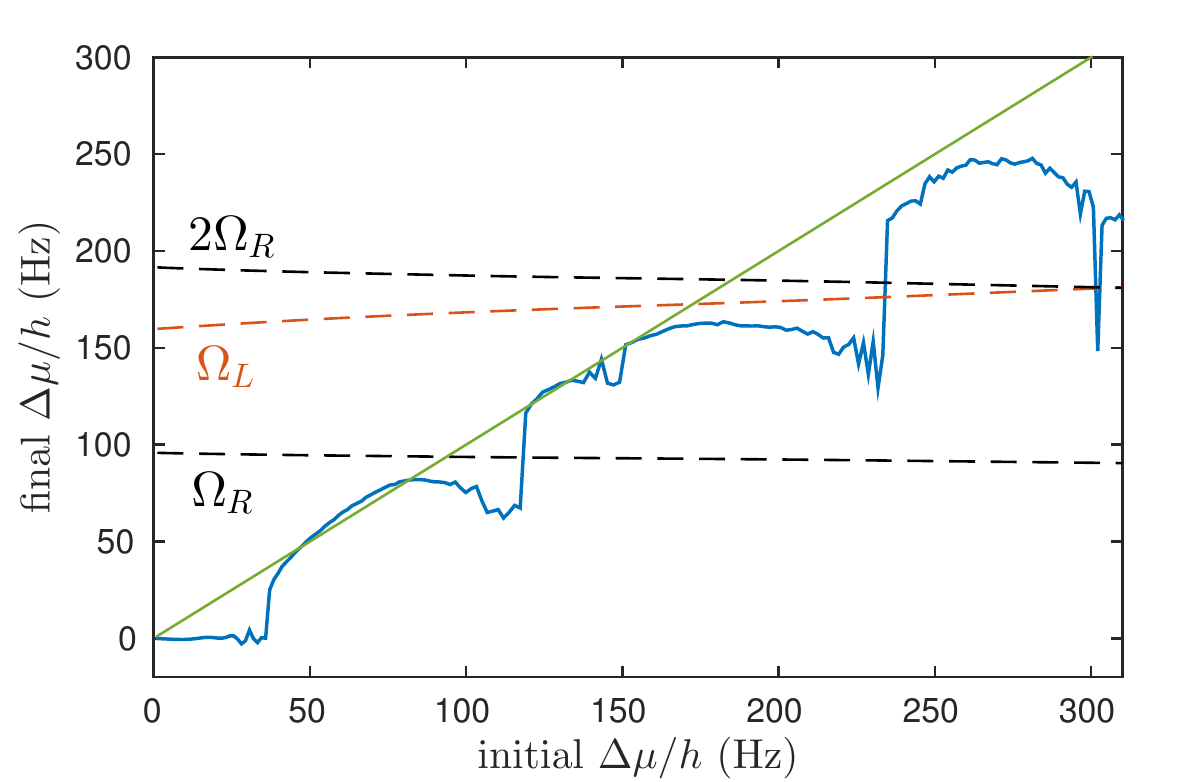}
\caption{\footnotesize
Phonon-Josephson resonances in an asymmetric system.
Shown on top corresponds to the barriers rotated by $\pi/8$ to the right ($\alpha_L = 5/4 \pi$, $\alpha_R = 3/4 \pi$).
The bottom corresponds to the barriers rotated by $\pi/8$ to the left  ($\alpha_L = 3/4 \pi$, $\alpha_R = 5/4 \pi$).
Notation and labeling are as in Fig.~\ref{fig:fiske}.
}\label{fig:pjres_shifted}
\end{figure}

In the symmetric trap discussed above the resonance frequencies in the left and right parts are very similar.
Therefore, we do not observe separate ``left'' or ``right'' resonances with phonon modes.
In order to separate left and right modes we change the angular positions of the barriers in such a way
that the angular extent of the left part is $\alpha_L = 3/4 \pi$ and that of the right part is $\alpha_R = 5/4 \pi$, or reversed. The right part always has lower density than the left one.
The results of these calculations are presented in Fig.~\ref{fig:pjres_shifted}.
In both cases the resonance regions are clearly observed, however, the positions of the minima of the chemical potential difference do not as accurately match the phonon frequencies as for symmetric barriers.
A possible reason for this discrepancy could be the considerable azimuthal inhomogeneity of the condensate parts in the case of the rotated barriers, while our estimates of the azimuthal sound wave frequencies are based on the assumption of  azimuthal homogeneity of the condensate.
This inhomogeneity can distort the sound waves in the condensate and also can produce additional excitations of different nature.
It can also be seen from Fig.~\ref{fig:pjres_shifted} that right resonances are considerably more pronounced then the left ones.

\subsection{Asymmetric height of the barriers}
\begin{figure}
\includegraphics[width=\linewidth]{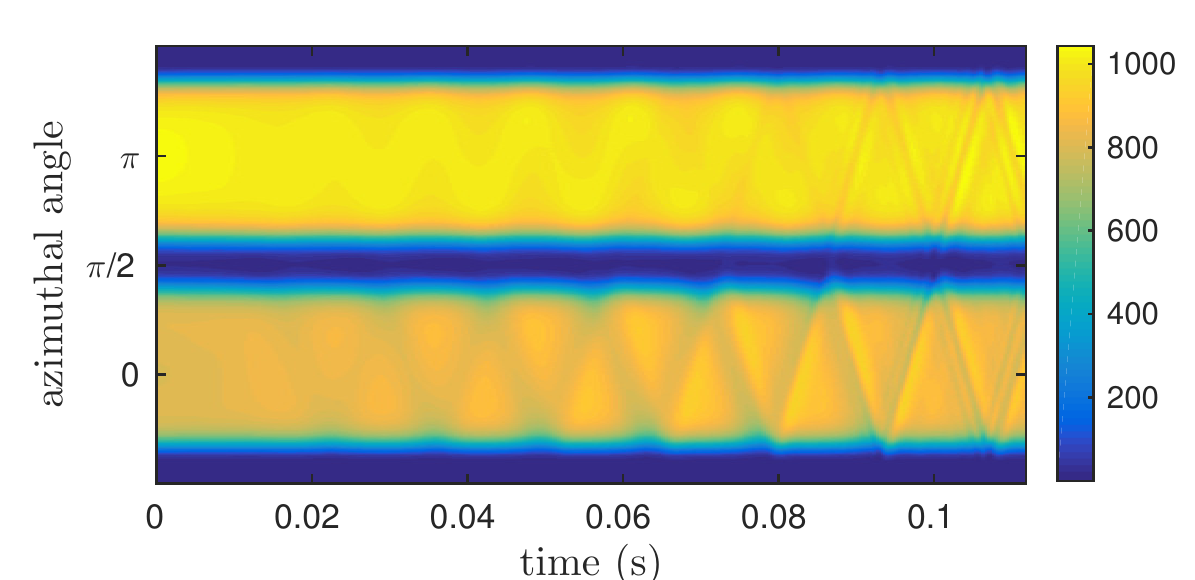}
\caption{Azimuthal distribution of the condensate density $n=|\Psi|^2$ as a function of time. The initial conditions correspond to the first observed resonance at $\Delta\mu/h \approx \Omega_{R,L}/2$.
}\label{fig:pjres_onejj_dens}
\end{figure}

We now analyze a setup with one of the barriers operating in the Josephson regime and the other one close to the Fock regime [$\omega_C/\omega_J \sim (N/2)^2$], i.e., tunneling is completely suppressed.
To this end we make a series of simulations with increased height of one of the barriers $U_1=4U_2$.
In this case we effectively have a system with only one Josephson junction as a source of resonant oscillations.
If the interaction between Josephson oscillations and sound modes only happens at one of the junctions, then the period of the lowest resonant phonon mode will not be the time that sound needs to travel from one Josephson barrier the other (as shown in Fig.~\ref{fig:phonon_az_dens}), but the time that it takes to travel from the only Josephson barrier back and forth. The phonon is reflected from the non-Josephson barrier (see Fig.~\ref{fig:pjres_onejj_dens}).
Therefore, we expect to observe an additional series of resonances at half-integer phonon frequencies.

\begin{figure}
\includegraphics[width=\linewidth]{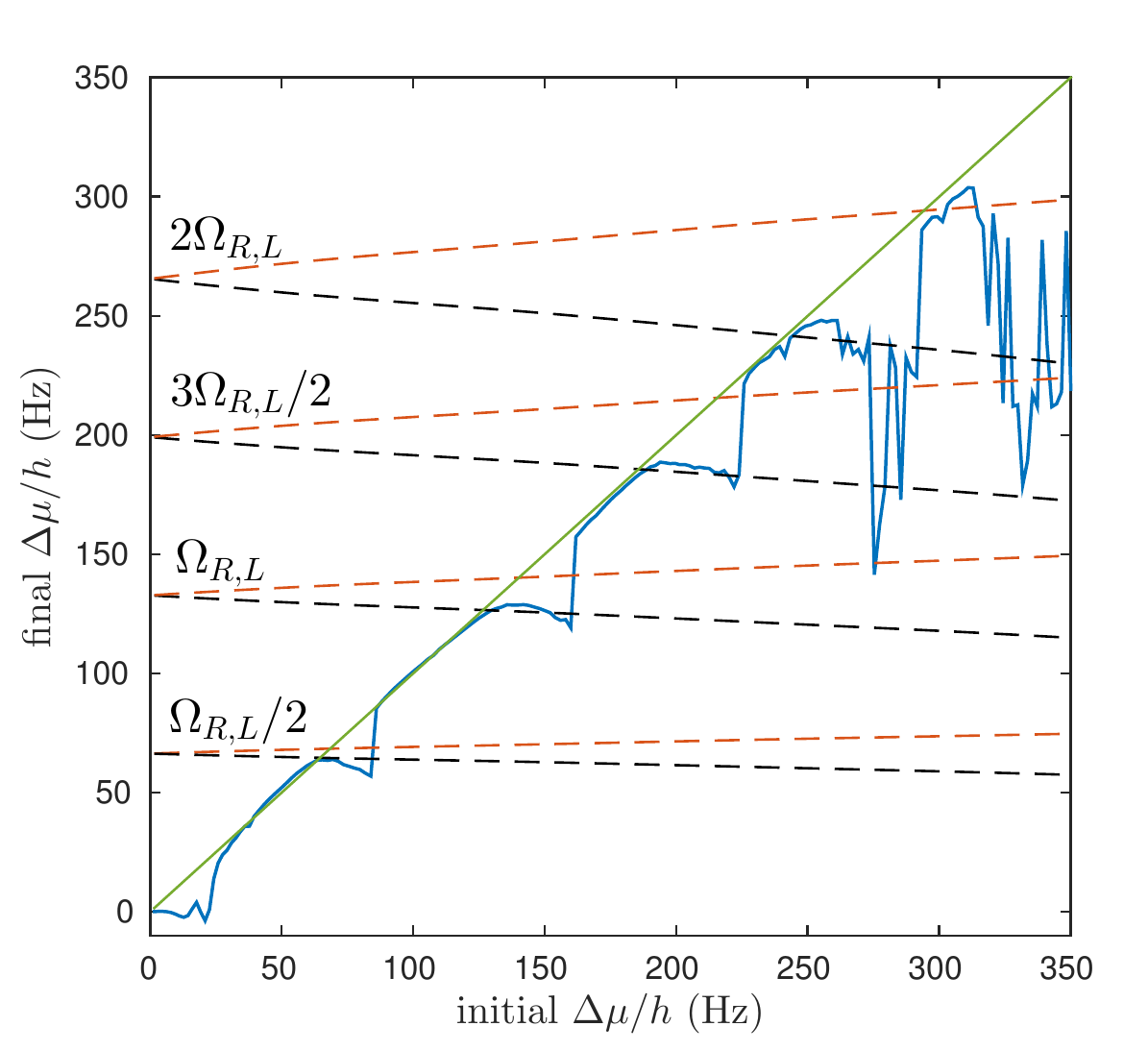}
\caption{Phonon-Josephson resonances if one barrier is outside the Josephson regime. Notation and labeling are as in Fig.~\ref{fig:fiske}.
}\label{fig:pjres_onejj}
\end{figure}

The result of this series of simulations is presented in Fig.~\ref{fig:pjres_onejj}.
We observe additional resonaces at half-integer phonon frequencies  as strong as the resonances at integer phonon frequencies.

The high barrier that we introduced  changes the topology of the system.
Such disconnected rings can be considered to be topologically equivalent to singly connected traps.
We therefore expect that a similar resonance picture is observed for a cigar-shaped condensate with just one Josephson junction.

\section{Conclusions}

In the present work we demonstrated that a coupling between phonon modes and Josephson oscillations in BEC can be observed in various traps with one or two Josephson barriers.
The resonant coupling manifests itself  as a reduction of the average population imbalance or chemical potential difference during the condensate evolution.
Alternatively, one may think of this effect as a phonon-induced additional tunneling current within the self-trapped Josephson regime.
It provides a dissipation channel even when thermal dissipation is suppressed.

The observed population imbalance shows a ladder like structure with higher-order resonances pronounced equally to the first one.
This indicates that the observation of Shapiro resonances in such systems may be strongly affected for driving frequencies higher then the lowest characteristic phonon frequency.

Our study suggests that Josephson oscillations may couple not only to sound modes but also to other low-energy collective modes (see also Ref.~\cite{Giovanazziphd}).
Such resonant couplings should be taken into account while engineering and interpreting BEC experiments with Josephson barriers.
The observed resonant coupling effect may be used for spectroscopy of phonons as well as for other low-energy collective excitations in Bose-Einstein condensates.
While the realization proposed here only distinguishes modes with well-separated frequencies, the resolution may be improved by fine-tuning the barrier parameters.

\begin{acknowledgments}
The authors are thankful to S. Eckel and A. I. Yakimenko for useful comments and discussions.
\end{acknowledgments}

\bibliography{RefsShapiro}
\end{document}